\documentclass[pre,aps, floatfix, superscriptaddress,showpacs]{revtex4-1}

\usepackage{graphicx}
\usepackage{dcolumn}
\usepackage{bm}

\newcommand{\be}{\begin{equation}} 
\newcommand{\ee}{\end{equation}}
\newcommand{\bea}{\begin{eqnarray}}   
\newcommand{\eea}{\end{eqnarray}}

\newcommand{\rr}{{\bf r}}
\newcommand{\qq}{{\bf q}}
\newcommand{\QQ}{{\bf Q}}
\newcommand{\eq}{{\bf e}_q}
\newcommand{\eQ}{{\bf e}_Q}

\newcommand{\eeta}{\boldsymbol{\eta}}
\newcommand{\bGamma}{\boldsymbol{\Gamma}}

\newcommand{\F}{\boldsymbol{F}}
\newcommand{\colxi}{\boldsymbol{u}}
\newcommand{\uu}{\boldsymbol{u}}

\newcommand{\OO}{{\cal O}}

\newcommand{\UU}{{\cal U}}

\usepackage[T1]{fontenc}
\usepackage[latin9]{inputenc}
\usepackage{amsmath}
\usepackage{amssymb}
\usepackage{graphicx}
\usepackage{esint}
\usepackage{lipsum}
\usepackage{mdframed}

\begin{document}

\preprint{AIP/123-QED}

\title{Pressure in an exactly solvable model of active fluid}

\author{Umberto Marini Bettolo Marconi}
\affiliation{Scuola di Scienze e Tecnologie, 
Universit\`a di Camerino, Via Madonna delle Carceri, 62032, Camerino, INFN Perugia, Italy}
\email{umberto.marinibettolo@unicam.it}
\author{Claudio Maggi}%
\affiliation
{ 
NANOTEC-CNR, Institute of Nanotechnology, Soft and Living Matter Laboratory, Piazzale A. Moro 2, I-00185, Roma, Italy }%
\author{Matteo Paoluzzi}
\affiliation{ Department of Physics, Syracuse University, Syracuse NY 13244, USA}

\date{\today}

\begin{abstract}
{ We consider the pressure in the steady-state regime of three stochastic models characterized by self-propulsion and persistent motion and widely employed to describe the behavior of active particles, namely the Active Brownian particle (ABP) model,  the
Gaussian colored noise (GCN) model and the unified colored noise model (UCNA). 
Whereas in the limit of short but finite persistence time the pressure in the UCNA model can be obtained by different methods which have an analog in equilibrium systems, in the remaining two models only the virial route is, in general, possible. 
 According to this method, notwithstanding each model obeys its own specific microscopic law of evolution, the pressure displays a certain universal behavior. For  generic 
 interparticle and confining potentials we derive a formula which establishes a correspondence between the GCN and the UCNA pressures.
In order to provide explicit formulas and examples, we specialize the discussion to the case of an assembly of elastic dumbbells confined to a parabolic well. By employing the UCNA we find that, for this model, the pressure determined by the thermodynamic method
coincides with the pressures obtained by the virial and mechanical methods. The three methods when applied to the GCN 
give a pressure identical to that obtained via the UCNA. Finally, we find that the ABP virial pressure 
 exactly agrees with the UCNA and GCN  result.}
%
\end{abstract}

\maketitle

\section{introduction}

The understanding of the properties of animated self-propelling agents is a challenge which recently attracted a profound interest in the condensed-matter physics community.
Such systems commonly referred to as "active matter'', are  able to convert energy from the environment into directed persistent motion either by metabolic processes,  as in the case of bacteria and spermatozoa, or by chemical reactions as in the case of
synthetic Janus particles \cite{marchetti2013hydrodynamics,bechinger2016active,maggi2016self,paoluzzi2016critical}.

From the theoretical point of view, it is interesting to study  the non-equilibrium steady-states (NESS), 
resulting from the balance between the energy continuously produced by the self-propulsion mechanism and the one  consumed  by dissipative forces exerted on the active particles by the viscous medium.
{ It is natural to ask whether one can characterize the NESS according  to few
macroscopic observables such as temperature, pressure and chemical potential and construct a "thermodynamic" theory.
A few years ago, Takatori et al.~\cite{takatori2014swim,takatori2015towards} discussed how to define and measure the pressure in active fluids and determined a new contribution to it stemming from the self-propulsion of the particles. 
They remarked that
an assembly of such particles as a result of their activity would swim away  unless confined by boundaries and identified the swimming pressure with the force per unit area necessary to constrain them inside that region of space.
The pressure problem has been recently tackled by statistical mechanics methods  by especially considering  two
 descriptions of active matter, namely the  active Brownian particle (ABP) model~\cite{solon2014pressure,joyeux2016pressure,speck2016ideal}}  and Gaussian colored noise model (GCN) \cite{szamel2014self,marconi2017heat,puglisi2017clausius} which are characterized by different modeling of the active driving force.
  In the case of confined spherical ABP with torque-free wall and interparticle forces,
 Solon and coworkers~\cite{solon2014pressure2} derived an expression for the mechanical pressure 
and proved that it is a state function
independent of the wall interaction, while
 Winkler et al. \cite{winkler2015virial}  considered a virial method  for ABP confined by solid walls or exposed to periodic boundary conditions. The pressure in the GCN model, instead, was recently studied by Fily et al. \cite{fily2016active}  and Sandford and Grosberg \cite{sandford2017memory}.
 In the case of the GCN, one can further simplify the analysis
by introducing a simplified model, the so-called unified colored noise approximation (UCNA)~\cite{hanggi1995colored,maggi2015multidimensional}.
A common ingredient to all these models is the presence of a persistence time $\tau$ which determines their 
special features, which do not have counterparts in equilibrium systems, such as the persistence of the  trajectories of the particles, correlated motions and decrease of their mobility as the density increases.
{ The  UCNA~\cite{marconi2016pressure} has the special property that its configurational 
 steady state distribution
 is known, and that its pressure 
 can be estimated by three different prescriptions 
 mutually consistent in the limit of small but finite $\tau$. These are:  a)  the  Clausius virial method~\cite{pathria1986statistical,widom2002statistical},
b)   a ``thermodynamic''  volume scaling method, which uses the volume derivative with respect to the system's volume of the partition function associated with the NESS distribution function 
and c) a method based on the evaluation of the work of deformation  in terms of the microscopic pair correlation function. }

{ Since exact results are scarce in this area, we consider instructive to tackle the pressure problem 
by applying different statistical methods to a minimal model for which, in many cases, as we are going to show,  the analysis can be performed without approximations.
The model was
introduced  by Riddell and Uhlenbeck  (RU) a long time ago \cite{riddell1950notion} to represent a collection of noninteracting  particle pairs mutually connected by harmonic springs (elastic dumbbells)
and confined by a harmonic trap. In spite of its simplicity, the RU model displays non trivial features, such as the
mobility reduction induced by the interactions, a dependence of the pressure on the persistence time $\tau$
and on the strength of the couplings.
    One of the advantages of the RU model is the possibility to compute the pressure exactly within the GCN and the UCNA
by the three methods above mentioned.
The explicit UCNA calculation shows that the different determinations of the pressure coincide to all orders in $\tau$ and not only to first order as one can prove for the case of a  generic potential. Interestingly, within the RU model such an equivalence between different  determinations of the pressure holds exactly also for the GCN and the GCN and UCNA pressures are identical. On the other hand,
concerning the ABP version of the RU model, we can only compute the pressure by the virial method and again we find that it has the same expression as the pressure of the GCN and UCNA models.}

{ 
The paper is organized as follows:
in section \ref{Models},
in order to allow the comparison between different treatments,  we first introduce 
the ABP, GCN, and UCNA models in the case of general interactions, establish the parameter correspondence among them and briefly review how to obtain the pressure in terms of steady-state statistical averages. 
In section \ref{RU},
we specialize the description and consider an application of our methods to the RU model.
 Finally, in section \ref{conclusions} we come to the conclusions and perspectives.  In order to limit the amount of mathematical details in the main text, we confined the technical aspects to four appendices.}

\section{Active particles models and their pressure}
\label{Models}

In the following, we consider the properties of an assembly of particles subjected to velocity dependent frictional forces  due to a solvent,  to conservative forces and to active forces.
Such a description has been employed in the recent literature and contains the minimal ingredients necessary to reproduce the basic features of an active fluid.  
The following equation describes the evolution of the positions of a set of particles suspended in an active bath 
\be
\dot \rr_{ i} =\frac{1}{\gamma}  \sum_{k}  \bGamma^{-1}_{i, k}  \Bigl[   {\bf F}_{ k}+ {\bf A}_{ k}
 \Bigl]
\label{stochasticxi}
\ee
where $i$ is the particle label and the matrix $\bGamma$ represents a non dimensional friction matrix whose particular form 
depends  on the model analyzed and will be specified below.
Each particle is driven by a conservative force  $\F_i= -\nabla_i \UU$, an active  force ${\bf A}_i$
 and  experiences a drag force,  $-\gamma \dot \rr_i$, with the solvent
 with drag coefficient $\gamma$.
${\bf A}_i$ is stochastic and correlated in time, $\tau$ being its characteristic time
associated with the
 persistent motion.
Equation \eqref{stochasticxi} is rather general and can represent
 few different models commonly used in active matter according to which
prescription is adopted for ${\bf A}_i$ and $\bGamma$.
We focus on three  models, namely the Active Brownian Particle (ABP) model \cite{romanczuk2012active,cates2013active,patch2017kinetics}, the Gaussian colored noise (GCN) \cite{fily2012athermal,farage2015effective}  or Active Ornstein-Uhlenbeck model
and the unified colored noise approximation (UCNA) model \cite{marconi2016effective}.
The ABP model was proposed on a phenomenological basis to describe
in terms of a set of stochastic differential equations
 containing a minimal set  of parameters  the observed behavior  of active fluids and somehow originated the remaining two models.
In fact, the GCN can be viewed as a simplified version of the ABP: it shares the same deterministic forces but is characterized by a Gaussian distribution
of the active force  ${\bf A}_i$ subject to the constraint of having the same variance and 
 time correlation as its counterpart in the ABP.
The UCNA, which is of interest because it lends itself to analytic treatments,  can be regarded as an (approximated)
reduction of the GCN to a Markovian form. 
Its dynamics  is local in time but is characterized by effective non pairwise forces 
not present in the GCN and stemming from the elimination procedure of the fast degrees of freedom present in the GCN.
The origin of these  non pairwise forces is the presence of non-diagonal terms in the UCNA matrix $\bGamma$
giving rise to an effective Hamiltonian
involving, in principle, two, three, four up to N-particles interactions.

Both in the ABP and GCN models the friction matrix $\bGamma$  has a very simple structure and  is the identity ${\bf I}$. 
 On the contrary,
the friction matrix $\bGamma$ is not diagonal  in the UCNA and
 contains in addition to ${\bf I}$ the Hessian matrix of the potential $\UU$,  i.e. $\bGamma={\bf I} +\tau {\bf H}$, where the elements
of   ${\bf H}$ are $-\frac{1}{\gamma} \frac{ \partial F_{\alpha i}  }{  \partial x_{\beta k}}$.
 In table \ref{table1},  for comparison, we report a synoptic view of the explicit form of the dynamic equation \eqref{stochasticxi}  in each model.
The deterministic force $\F_i$ is assumed to be the same in each model. 
{ The third key ingredient is the active force ${\bf A}_i=\gamma v_0 {\bf e}_i$.
As also illustrated in  table  \ref{table1} the active force ${\bf A}_i$ in the ABP is modeled by a vector of  constant intensity $\gamma v_0$, where $v_0$ is the propulsion speed,  and random orientation,
${\bf e}_i$, performing a  diffusive motion,
on the unit sphere (or on the unit circle in $d=2$) with rotational 
 diffusivity constant $D_r$.}
 The study of such a process involves the
probability distribution not only of the positions, but also of the angles.  The ABP micro-state, $\mu$,  in three dimensions
is specified by $N$ positions and $2N$ angles (  $\mu =\{\rr_i,\theta_i,\phi_i\}$ ) and  the angular dynamics is
 \bea
 \dot \theta_i(t) &=& D_r \cot \theta_i + \sqrt{D_r} \eta^{\theta}_i   \nonumber
 \\
\dot \phi_i(t)&=&   \frac{\sqrt{D_r}}{\sin\theta_i} \eta^{\phi}_i(t)  \, .
\label{angulard3}
\eea
In two dimensions instead  one needs only $N$ angles $\theta_i$ evolving as:
  \be
\dot \theta_i(t) = \sqrt{D_r} \eta^{\theta}_i(t)
\label{angulard2}
\ee
and the micro-state is  $\mu =\{\rr_i,\theta_i\}$. The noises have zero average and correlations
 $
\langle \eta^{\theta}_i(t) \eta^{\theta}_j(t')\rangle=2 \delta_{ij} \delta(t-t') \, .
$
{ If one uses the orientations, instead of the angles, it is easy to show that ${\bf e}_i$  at two different instants $t$ and $t'$ are exponentially correlated
\be
\langle  {\bf e}_i(t) {\bf e}_j(t')  \rangle = e^{-(d-1) D_r |t-t'|}\boldsymbol{1}\delta_{ij} \, .
\label{eiej}
\ee }

The Gaussian colored noise model (GCN) represents a convenient alternative to the ABP and  has been introduced with the main motivation of  simplifying the theoretical study
of active fluids,  because it releases the hard constraint  of constant magnitude of the velocity.
{   The idea is to eliminate  the active ABP force, $\gamma v_0 {\bf e}_i$, in favor of a 
non-Markovian term $\gamma u_i$, where the variable $u_i$ represents an Ornstein-Uhlenbeck process of characteristic time $\tau$
 and whose governing equation is:
 \be
 \dot\colxi_i(t) =- \frac{1}{\tau}\colxi_i(t) + \frac{ D^{1/2}}{ \tau} \eeta_i(t) \, .
 \ee 
 }
In the
 GCN each component of $\uu_i$ is allowed to fluctuate between $-\infty$ and $\infty$, according to a Gaussian distribution of zero average and variance $D/\tau$.
By an appropriate choice of the parameters
$\tau$ and $D$ one can establish a correspondence between the ABP and the GCN statistical properties.
In the GCN, the diffusion coefficient $D$ is related to the original parameters by $D=\frac{v_0^2 }{d(d-1) D_r}  $,
while $\tau$ the persistence time is $\tau^{-1}=(d-1)D_r$ while
the time self-correlation of the velocity is by construction  exponentially decaying as
eq.~\eqref{eiej} and adjusted to 
reproduce the velocity self-correlation of the ABP.  The GCN micro-state, $\mu=\{ \rr_i,\uu_i\} $,  of the system is 
the set of the positions and of the velocities of each particle.

Finally, the UCNA represents a further approximation, aimed to simplify the analytic work, and may be derived from the
GCN: it is tantamount of performing an adiabatic approximation replacing the colored noise term by an effective Markovian white noise. Such an elimination procedure brings about a complicated structure of the equations of evolution for the particles coordinates, due to the presence of a friction matrix $\bGamma$ which in principle couples the motion of all  particles \cite{maggi2015multidimensional,marconi2015towards,marconi2016effective}.

{  In the UCNA case, 
 the evolution equation \eqref{stochasticxi} of the particle positions $\rr_1,\dots,\rr_N$, identifying the micro state $\mu$,  
 takes the following specific form:
\be
\dot r_{\alpha i} \simeq \sum_{\beta k} \Gamma^{-1}_{\alpha i,\beta k}  \Bigl[ \frac{1}{\gamma}  F_{\beta k}+
D^{1/2}  \eta_{\beta k}(t)   \Bigl] .
\label{stochasticxib}
\ee
with
\be
\Gamma_{\alpha i,\beta k}=\delta_{ik}\delta_{\alpha \beta} +\frac{\tau}{\gamma} \frac{\partial^2 \UU }{\partial x_{\alpha i}  \partial x_{\beta k}}
=\delta_{ik}\delta_{\alpha \beta} -\frac{\tau}{\gamma} \frac{ \partial F_{\alpha i}  }{  \partial x_{\beta k}}.\label{gammamatrix}
\ee
where Greek indexes stand for Cartesian components. 
The stochastic terms  $\eeta_i(t)$ are Gaussian and Markovian processes distributed with zero mean
and moments $
\langle \eeta_i(t) \eeta_j(t')\rangle=2 \boldsymbol{1} \delta_{ij}  \delta(t-t')
$.  
Equation \eqref{stochasticxib}
 shows that  the effective friction and the effective noise
experienced by each particle 
depend on the coordinates of all other particles.  }


\begin{table*} 
\begin{tabular}{ |p{1cm} |p{2.9cm}|p{2.3cm}|p{4.5cm}|p{5.1cm}| p{2.5cm}| }
\hline
\multicolumn{6}{|c|}  {Stochastic   dynamics of different active models}\\
\hline
Model &$   \bGamma_{i, k}  $  & $ {\bf A}_{ i}$& Active dynamics& Active noise Correlations&Diffusivity \\
\hline
ABP3& ${\bf I}_{i, k}$&  $\gamma v_0 {\bf e}_i(t)$  & $\dot \theta_i(t) =D_r \cot \theta_i + \sqrt{D_r} \eta^{\theta}_i  $\,\,
\,  $\dot \phi_i(t)=   \frac{\sqrt{D_r}}{\sin\theta_i} \eta^{\phi}_i(t) $&$
\langle  {\bf e}_i(t) {\bf e}_j(t')  \rangle = e^{-2 D_r |t-t'|}\boldsymbol{1}\delta_{ij}$& $2 D_r=1/\tau$ \\
\hline
ABP2& ${\bf I}_{i, k}$&  $\gamma v_0 {\bf e}_i(t)$  & $\dot \theta_i(t) = \sqrt{D_r} \eta^{\theta}_i(t)$&$
\langle  {\bf e}_i(t) \cdot{\bf e}_j(t')  \rangle = e^{-D_r |t-t'|}\delta_{ij}$& $ D_r=1/\tau$ \\
\hline
GCN& ${\bf I}_{i, k}$& $\gamma\uu_i(t)$   & $\dot\colxi_i(t) =- \frac{1}{\tau}\colxi_i(t) + \frac{ D^{1/2}}{ \tau} \eeta_i(t) $ &$\langle\colxi_i(t)\,\colxi_j(t')\rangle =  \frac{D}{\tau} e^{-|t-t'|/\tau}\boldsymbol{1}\delta_{ij}$ &$D=\frac{v_0^2 }{d(d-1) D_r}  $ \ \\
\hline
UCNA & $\delta_{ik}\delta_{\alpha \beta} +\frac{\tau}{\gamma} \frac{ \partial^2 \UU  }{ \partial x_{\alpha i}  \partial x_{\beta k}    }$& $\gamma D^{1/2}  \eeta_{ i}(t)$  & $\eeta_i(t)$ is a Wiener process &$ \langle \eeta_{ i}(t)   \eeta_{ j}(t') \rangle =2  {\boldsymbol\delta}_{ij} \delta(t-t')$&D\\
\hline
\end{tabular}
\caption{Comparison between the stochastic dynamics describing the ABP in 3 and 2 dimensions, the GCN  and the UCNA. For each model, in the second column we report the effective friction matrix 
$\bGamma$, in the third column the active force ${\bf A}$, in the fourth we specify the stochastic equation associated with the active driving, in the fifth column we display the noise correlations and in the sixth the diffusivity of each model.}
\label{table1}
\end{table*}

\vspace{0.5cm}
\subsection{Virial pressure}
We turn now to review the methods to determine the pressure in the above models.
Since the methods are based on the knowledge of the NESS distribution functions,
instead of using the stochastic differential equation \eqref{stochasticxi}, it is more convenient
to use the associated Fokker-Planck equation (FPE) describing the evolution of the  probability distribution function of the micro states, $\mu$. For all cases studied in the present paper the FPE can be written as:
\be
\frac{\partial }{\partial t} P(\mu,t)={\cal L}_{FP} P(\mu,t)
\label{fpmu}
\ee 
where the specific form of the Fokker-Planck operator ${\cal L}_{FP}$,  the sum of a diffusive and a drift   contribution, is given
explicitly in the following.
The average of an observable $\OO(\mu)$ evolves according to
\be
\frac{\partial }{\partial t}\langle \OO(\mu,t)\rangle =\langle{\cal L}^{\dagger}_{FP} \OO(\mu,t)\rangle
\ee 
where ${\cal L}^{\dagger}_{FP} $ is the adjoint operator of the Fokker-Planck operator ${\cal L}_{FP} $ and
$\langle \OO(\mu,t)\rangle \equiv\int d\mu P(\mu,t) \OO(\mu)$.


\subsubsection{Virial  pressure in the ABP and GCN models}

To determine the pressure we employ the virial method ~\cite{falasco2015mesoscopic,winkler2015virial} to obtain the virial of the forces.
In $d=3$ the  ABP evolution equation for a generic operator $\OO(\rr_i,\theta_i,\phi_i)$ reads:
\begin{widetext}
\bea
\frac{d  \langle\OO(t) \rangle^{ABP}}{dt }&=&  \int d^N \rr  \int_0^{2\pi} d^N\phi  \int_0^\pi d^N \theta \sin\theta \, P^{ABP}_N(\rr_i,\theta_i,\phi_i,t) \,
\sum_{ i}  \Bigl(\frac{1}{\gamma }  (\F_i+ {\bf A}_i)  \frac{\partial }{\partial \rr_i} \,    
  \nonumber \\
&&
+ D_r  (  \frac{1}{\sin \theta_i} \frac{\partial}{\partial \theta_i} \sin\theta_i  \frac{\partial}{\partial \theta_i}  + \frac{1}{\sin^2\theta_i}   \frac{\partial^2 }{\partial \phi_i^2}   ) \Bigr) \OO(\rr_i,\theta_i,\phi_i) 
\label{operator3dim}
\eea
\end{widetext}
{ with ${\bf A}_i=\gamma v_0 {\bf e}_i$}.

On the other hand,
in the GCN model the evolution equation for $\OO(\rr_i,\uu_i)$   is:
\begin{widetext}
\be
\frac{d \langle\OO(t)  \rangle^{GCN}}{dt}=  \int d^{N} \rr   \int d^{N} \uu P_N^{GCN}( \rr_i,\uu_i,t)
\sum_i  \Bigl(\frac{1}{\gamma }  (\F_i+{\bf A}_i)  \frac{\partial }{\partial \rr_i}  \,     -\frac{  \uu_i }{\tau}    \frac{\partial  }{\partial \uu_i}  +  \frac{D}{\tau^2}   \frac{\partial^2 }{\partial \uu_i^2}          \Bigr) \OO(\rr_i,\uu_i)
\label{oevolution1}
\ee
\end{widetext}
{ and ${\bf A}_i=\gamma {\bf u}_i$}.
The virial of the forces is obtained by
choosing
$\OO=\sum_i \rr_i\cdot \rr_i$, that is the mean square displacement of the particles  with respect to the origin.
 Its average is asymptotically bounded by the presence of the walls and its derivative vanishes as for $t\to \infty$. 
 In both models, one obtains the following equation, relating the virial of the 
external forces      $- p_v  d L^d=\sum_i^N  \langle \F^{ext}_i \cdot \rr_i \rangle $   to the internal virial $\frac{1}{2}\sum_{i\neq j }  \langle \F_{i j} \cdot (\rr_i-\rr_j) \rangle$ and to the average of the moment arm of the active force    $ \langle {\bf A}_i\cdot\rr_i \rangle$:
\be
 p_v V L^d=  \frac{1}{2}\sum_{i\neq j} \langle \F_{ij} \cdot (\rr_i-\rr_j) \rangle+
 \sum_i \langle {\bf A}_i\cdot\rr_i \rangle .
\label{virialbaiesi1}
\ee
Eq.~\eqref{virialbaiesi1}  is not a closed equation because  the ABP and GCN  not only require the knowledge of the probability distribution of the particle  positions,
but also of the correlations between these and the  active forces ${\bf A}_i$. Thus, we need to consider the evolution of the average of the operators   $\OO= \rr_i \cdot  {\bf e}_i $ and    $\OO= \rr_i \cdot  \uu_i$,  in the case of the ABP and GCN, respectively. In both instances, we can write an auxiliary equation
under the same form:
\bea
& & \frac{ \gamma}{\tau}  \sum_i  \langle   {\bf A}_i \cdot  \rr_i  \rangle = 
N d D \frac{\gamma^2}{\tau}    +\sum_i  \langle \F^{ext}_i \cdot {\bf A}_i \rangle +
\nonumber \\
& & \frac{1}{2}\sum_{i\neq j} \langle \F_{ij} \cdot ({\bf A}_i-{\bf A}_j) \rangle \, ,
\label{virialbaiesi88}
\eea
{ where we have used the correspondence relations $\tau^{-1}=(d-1) D_r $  and $D=\frac{v_0^2 }{d(d-1) D_r}  $
of table I in order to stress the similarity.
The expression for the ABP and GCN pressure is  obtained from  \eqref{virialbaiesi1}, where  the r.h.s. represents the internal pressure, whose form is reported for each case in the second column of  table \ref{table3}, whereas the third column reports the corresponding form of the auxiliary equation \eqref{virialbaiesi88}.
}
In the GCN case, we may derive in the small $\tau$ limit a closed equation for the steady-state average $\langle {\bf A}_i\cdot\rr_i \rangle $, as shown in appendix~\ref{equivalence}, whereas in the ABP we could obtain
an explicit result only for a specific model comprising harmonic forces
(see  subsection~\ref{virialABP}).
\begin{table*}
\begin{tabular}{ |p{1cm} |p{7.5cm}|p{10cm}| }
\hline
\multicolumn{3}{|c|}{Virial pressure} \\
\hline
Model &  Internal pressure $p_v d L^d  $   & Auxiliary equation  \\
\hline
ABP3  & $\frac{1}{2}\sum_{ij}' \langle \F_{ij} \cdot (\rr_i-\rr_j) \rangle+
\gamma  v_0 \sum_i \langle {\bf e}_i \cdot\rr_i \rangle$ &$  \sum_i  \langle   {\bf e}_i \cdot  \rr_i  \rangle = N \tau v_0 + \frac{\tau}{\gamma}[\sum_i  \langle \F^{ext}_i \cdot {\bf e}_i \rangle +\frac{1}{2}  \sum_{ij}' \langle \F_{ij} \cdot ({\bf e}_i-{\bf e}_j) \rangle]
$\\
ABP2&  $ \frac{1}{2}\sum_{ij}' \langle \F_{ij} \cdot (\rr_i-\rr_j) \rangle+
\gamma  v_0 \sum_i \langle {\bf e}_i\cdot\rr_i \rangle $ & $  \sum_i  \langle   {\bf e}_i \cdot  \rr_i  \rangle = N \tau v_0 +
 \frac{\tau}{\gamma}\sum_i [ \langle \F^{ext}_i \cdot {\bf e}_i \rangle +\frac{1}{2}\sum_{ij}' \langle \F_{ij} \cdot ({\bf e}_i-{\bf e}_j) \rangle]
$\\
GCN  & $  \frac{1}{2}\sum_{ij}  \langle \F_{ij} \cdot (\rr_i-\rr_j) \rangle+
\gamma  \sum_i \langle \uu_i\cdot\rr_i \rangle$ & $   \sum_i  \langle   \uu_i\cdot\rr_i\rangle=
 N d D + \frac{\tau}{\gamma} [\sum_i  \langle \F^{ext}_i \cdot \uu_i \rangle + \frac{1}{2} \sum_{ij}'  \langle \F_{ij} \cdot ( \uu_i-\uu_j) \rangle]
$  \\
UCNA & $
 \frac{1}{2 }\sum_{ij}'  \langle \F_{ij} \cdot (\rr_i-\rr_j) \rangle +  D\gamma \sum_{\alpha i} \langle   \Gamma^{-1}_{\alpha i,\alpha i}  \rangle$& $\sum_{\alpha i}  \langle   \Gamma^{-1}_{\alpha i,\alpha i}  \rangle= \langle Tr [{\bf I}+\tau {\bf H}    ]^{-1}\rangle $ \\
\hline
\end{tabular}
\vspace{0.5cm}
\caption{ For each model, in the second column we display the contribution to the pressure due to the internal forces and to the diffusive dynamics. In the third column we report the auxiliary equation (given in the main text as eq.~\eqref{virialbaiesi88} for ABP and GCN, and 
as eq.~\eqref{b2} for the UCNA)
needed to obtain a closed expression for the pressure.}
\label{table3}
\end{table*}

\subsubsection{Virial  pressure in the UCNA}
{ In the case of general potentials,  the UCNA formulas for the  pressure were derived in detail in ref.~\cite{marconi2016pressure}:
the virial pressure was shown to have the form:
\be
p_v V d=  
 \frac{1}{2}\sum_{i \neq j}  \langle \F_{ij} \cdot (\rr_i-\rr_j) \rangle 
+  D\gamma \sum_{\alpha i} \langle   \Gamma^{-1}_{\alpha i ,\alpha i}  \rangle
  \label{virialbaiesivv}
\ee
with $\sum_j \F_{ij}\equiv\sum_j \F^{particles}(\rr_i-\rr_j)= \F^{int}(\rr_i)$ and 
 $  \Gamma^{-1}_{\alpha i,\beta j}  =\Bigl[\delta_{ij} \delta_{\alpha \beta}- \frac{\tau}{\gamma }  
 \frac{\partial F^{ext}_{\alpha i} }{\partial r_{\beta j} }- \frac{\tau}{\gamma }   \frac{\partial  F^{int}_{\alpha i} }  {\partial r_{\beta j}  }   \Bigr ]^{-1}$.
 The first term in eq. \eqref{virialbaiesivv} is the analog of the
 non ideal pressure contribution in a passive fluid, whereas  the second term represents the swim pressure.}

{ One may observe the similarity between  eqs.   \eqref{virialbaiesi1} and \eqref{virialbaiesivv}:  the pressure is given by a direct interaction term involving 
pair forces and analogous to the contribution to the pressure of passive fluids stemming from the pair potential plus a term due to the
presence of  active forces.
In appendix \ref{equivalence}, we shall show that the second terms featuring in the GCN \eqref{virialbaiesi1} and UCNA \eqref{virialbaiesivv}   pressure equations
 are equivalent to first order in $\tau$.}

\subsection{Thermodynamic and Mechanical determinations of the pressure}

{ To first order in the non equilibrium parameter $\tau$,
two alternative procedures to measure the pressure in active systems
\cite{marconi2016pressure} may also be applied:}
the first of these is  a "thermodynamic" method. Let us suppose that we are able to determine the non-equilibrium
steady-state distribution of micro-states, $\mu$, and its normalizing factor, $Z$.
We may identify $Z$ with the relevant  partition function and the pressure
with its logarithmic
derivative  with respect to the volume times the temperature. In reference \cite{marconi2016pressure}  it was verified that within the UCNA up to
first order in $\tau$ such a "thermodynamic" pressure coincides with 
 the virial  pressure. 

The second alternative procedure consists of a mechanical  method to measure the pressure and employs the concept of work involved to increase the volume of the system and 
relates the force necessary to perform it  to the microscopic structure of the system~\cite{baus1991stress,lovett1991family}. One takes advantage of the fact that
the mechanical work associated with a strain,   resulting from  a nonuniform displacement of the particles of the fluid,  can be 
expressed either in terms of the external force field which produces it or in terms of the product of the stress and strain tensors.
 Using the information contained in the microscopic distribution one and two-particles distribution functions one can determine the
 mechanical work and finally the pressure.

 \section{ Application to the Riddell-Uhlenbeck active particles model}
 \label{RU}
 In general, for given confining and inter-particle potentials  it is not possible to write explicitly the pressure
 in terms of its control variables and to compare its expressions derived by different methods in order to verify
 their compatibility. As we discussed above the comparison shows an agreement to first order in $\tau$,
 but it is difficult to go beyond for an arbitrary choice of potentials.
  In the following, in order to compare the pressure of different  methods and different descriptions of the active forces,   
 we shall employ an explicitly  soluble model. The system is represented by $N$  mutually noninteracting elastic dumbbells, i.e. two point particles bound together by
 an elastic spring of constant $\alpha^2$, moving in a vessel represented by a harmonic weak confining  potential, of spring constant  $\omega^2$. 
  Such a model, similar to the harmonic trap model \cite{szamel2014self,maggi2014generalized,takatori2016acoustic}, was proposed long ago by  Riddell and Uhlenbeck. It contains the minimal ingredients to observe the competition between 
 internal forces and confining potential and can be solved without introducing further approximations.
  The  RU \cite{riddell1950notion} 
 potential energy  reads:
$$\UU(\rr_1,\rr_2)=w(\rr_1-\rr_2)+u(\rr_1)  +u(\rr_2) $$
with $w(\rr)=\frac{1}{2}\alpha^2 \rr^2$. By setting
$u(\rr)=\frac{k}{2}\frac{  \rr^2 }{ L^2}$, one  introduces a volume dependence in the spring constant associated with the confining potential and for simplicity of notation we shall use $\omega^2=\frac{  k}{ L^2}$.
 Provided the condition $-\alpha^2<\omega^2/2$ is satisfied,
it is also possible to include the case of  repulsive inter-particle quadratic potentials, which have been used in simulations
of active particles \cite{patch2017kinetics}.
We follow this strategy: for each of the three models, we derive, when possible, the pressure formula.
In the UCNA and  GCN  cases, we  apply the three routes and verify that the results of each case coincide, whereas in the ABP case we are able to obtain the
pressure only via the virial route, but we show that the pressure is  the same as the other six cases.

\subsection{UCNA analysis of the Riddell-Uhlenbeck active model}

By the adiabatic elimination of the velocities we obtain 
the corresponding  approximate UCNA equations \eqref{stochasticxi} . These equations are more conveniently written
in terms of the "collective" coordinates $\qq=(\rr_1-\rr_2)$ and $\QQ=(\rr_1+\rr_2)/2$  and of the renormalised  spring constant $\Omega^2=\omega^2+2\alpha^2$
as:
\bea
 \dot \qq&=&-  \frac{1} {1+\frac{\tau}{\gamma} \Omega^2}  \Bigl( \frac{1}{\gamma}\Omega^2 \qq+D^{1/2}(\eeta_1-\eta_2) \Bigr)\\
 \dot \QQ&=&- \frac{1} {1+\frac{\tau}{\gamma} \omega^2} \Bigl(\frac{1}{\gamma} \omega^2 \QQ+D^{1/2} \frac{(\eeta_1+\eeta_2)}{2} \Bigr)
\eea
\subsubsection{Thermodynamic route to the pressure}

\begin{figure}[!t]
\centering
\includegraphics[width=.5\textwidth]{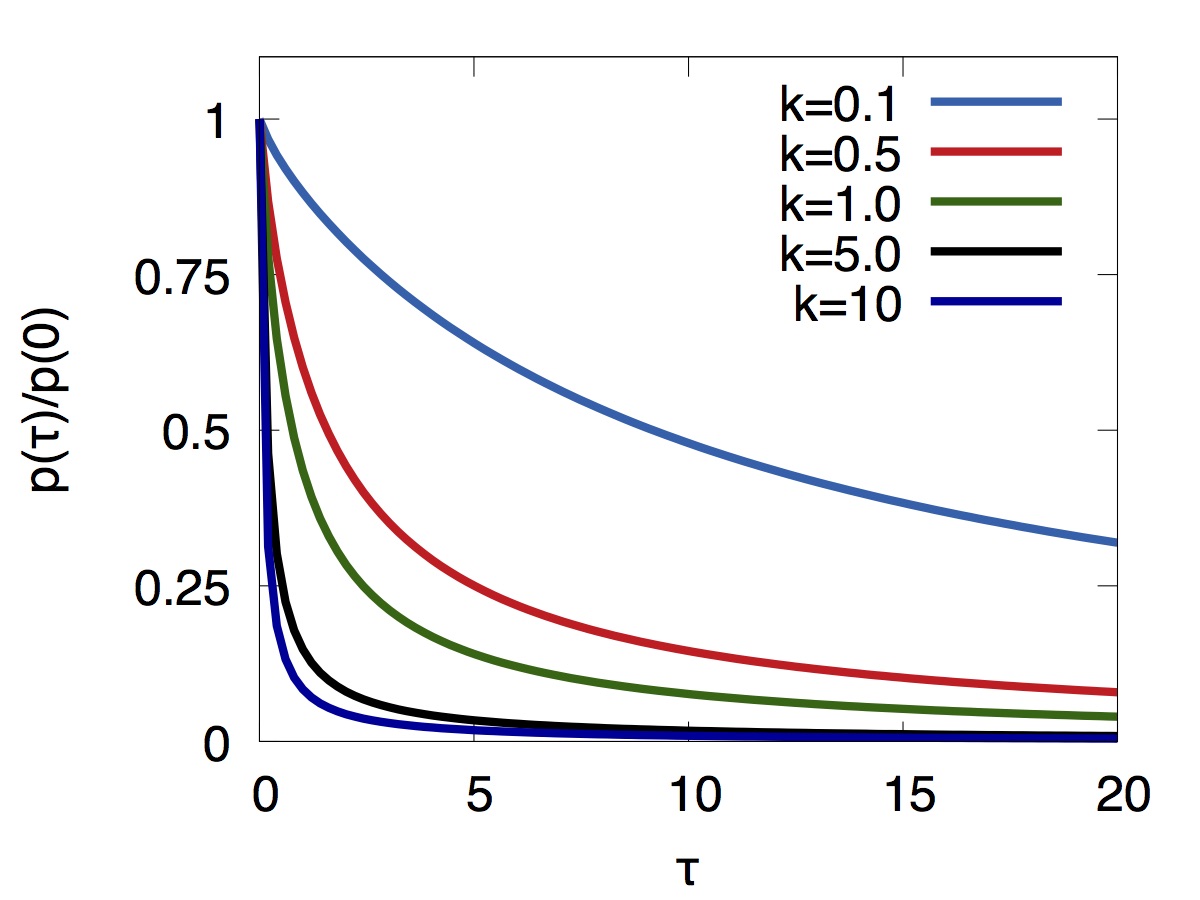}
\caption{Pressure as a function of $\tau$ for different values of $k=0.1,0.5,1.0,5.0,10$ (blue,red,green,black, and dark blue, respectively) and $L=\gamma=\alpha=1$.}
\label{fig:pressure}      
\end{figure}

It is quite straightforward to obtain the steady-state non-equilibrium distribution function of the  UCNA model.  It reads:
\bea
& & P(\qq,\QQ)= 
\nonumber \\ 
& & \frac{1}{Z}
\exp\Bigl\{- \beta [\omega^2(1+\frac{\tau}{\gamma}\omega^2)  \QQ^2+\frac{1}{4}  \Omega^2(1+ \frac{\tau}{\gamma}\Omega^2 )
\qq^2] \Bigr\}  \, \det \Gamma \nonumber \\
\eea
where we used the abbreviations  $1/\beta=D\gamma$ and
$\det \Gamma=(1+\frac{\tau}{\gamma} \omega^2)^d (1+\frac{\tau}{\gamma} \Omega^2)^d$.
Integrating over $\qq$ and $\QQ$ we obtain the "partition function" of the UCNA model:
 $
 Z= 2\pi\frac{\Bigl( (1+\frac{\tau}{\gamma}\omega^2 ) (1+\frac{\tau}{\gamma}\Omega^2)   \Bigr)^{d/2} }{ (\beta \omega \Omega)^d} 
 \label{zucna}
 $
 and identify  the logarithmic  volume derivative of the partition function with a thermodynamic pressure :
  \be
 p_t=  \frac{1}{\beta}\frac{\partial }{\partial L^d} \ln Z= 
  \frac{ D\gamma }{ L^d} \Bigl\{  
\frac{1}   {1+\frac{\tau}\gamma\omega^2}+ \frac{ \omega^2 }{ \Omega^2} 
\frac{1}   {1+\frac{\tau}\gamma\Omega^2} \Bigr\}
\label{Upt}
\ee
In Fig. \ref{fig:pressure} we show the behavior of $p_t$ given by equation (\ref{Upt}) as a function of $\tau$. 
We normalized $p_t(\tau)$ with the value $p(0)=p_{eq}$, where $p_{eq}$ it the equilibrium value
of the pressure of an ideal  gas of elastic
dumbbells.
The different curves refer to different values of $k=0.1,0.5,1.0,5,10$. The remaining parameters are fixed to one, i. e., $L\!=\!\gamma\!=\!\alpha\!=\!D\!=\!1$. As one can see the pressure monotonically decreases with $\tau$ and $\lim_{\tau\to\infty}p_t=0$
 when $D$ is kept fixed.
%

\subsubsection{ Clausius Virial pressure in the UCNA model } 
The virial pressure is obtained by applying the general formula \eqref{virialbaiesivv}  with the choice $\OO=\frac{1}{2}
\sum_{ij}\sum_{\alpha}\Gamma_{i j} r_{\alpha i}r_{\alpha j}$:
\bea
& &
 p_v d L^d = 
 -\langle \F^{ext}_1 \cdot \rr_1+\F^{ext}_2 \cdot \rr_2\rangle= 
  \nonumber \\
 & &
 d D\gamma(\Gamma^{-1}_{11}+\Gamma ^{-1}_{22}) +\langle \F_{12} \cdot (\rr_1-\rr_2)\rangle .
 \eea
 Using the results $\Gamma^{-1}_{11}+\Gamma ^{-1}_{22} =\frac{1}   {1+\frac{\tau}\gamma\omega^2}+\frac{1}   {1+\frac{\tau}\gamma\Omega^2}$ and  $\langle \F_{12} \cdot (\rr_1-\rr_2)\rangle =  \frac{ D\gamma}{ \Omega^2}  \frac{\omega^2-\Omega^2}   {1+\frac{\tau}\gamma\Omega^2}$ we  find $p_v=p_t$
and  conclude that the mechanical pressure and the thermodynamic pressure are equal for this model.
Notice that as $\alpha\to \infty$ the pressure of the dumbbell gas is only one-half of the pressure   of a system of noninteracting particles
contained in the same vessel, since each dumbbell behaves as a single particle.




 \subsubsection{Mechanical pressure  in the UCNA model }
We have recently shown \cite{marconi2015towards} that
 within the UCNA it is possible to derive an exact hierarchy of equations, similar to the Born-Green-Yvon   equations, connecting the distribution function, $P_N^{(m)}(\rr_1,\dots,\rr_m)$, of $m$ particles, in systems containing $N$ of them, to those of $m' >  m$ particles.
As shown in appendix \ref{mechanicalRU} we find that within the UCNA   the virial, thermodynamic and distribution function route to computing the  pressure  give the same result, i.e. $p_v=p_t=p_V$
as in the case of equilibrium systems. 

 \subsection{  Riddell-Uhlenbeck  model with Gaussian colored noise and for Active Brownian particles} 
We turn, now, to the study of the pressure  in the RU model when the dynamics  follow the GCN or the ABP
prescription.
The coordinates of the two particles evolve according to the  following equations :
\bea
 \dot \rr_1&=&-\frac{1}{\gamma} \Bigl(\omega^2 \rr_1+\alpha^2 (\rr_1-\rr_2)-{\bf A}_1(t) \Bigr)
 \label{x1dot} \\
 \dot \rr_2&=&-\frac{1}{\gamma}\Bigl( \omega^2 \rr_2-\alpha^2 (\rr_1-\rr_2)-{\bf A}_2(t) \Bigr)
\label{x2dot}
\eea
The GCN  corresponds to the choice ${\bf A}_i=\gamma \uu_i(t)$ 
  with:
$
\dot \uu_i(t) =- \frac{1}{\tau} \uu_i(t) + \frac{ D^{1/2}}{ \tau} \eeta_i(t) 
$, whereas the ABP to ${\bf A}_i=\gamma v_0 {\bf e}_i(t)$,
 where the unit vectors ${\bf e}_i$ identified by their angles evolve according to \eqref{angulard3} and \eqref{angulard2}
 in three  and two dimensions, respectively.
\subsubsection{RU+GCN model: virial route}

We consider first the GCN case and derive the pressure by the virial and thermodynamic method.
We show that within the  GCN  the virial expression 
of the pressure gives the same result as   \eqref{Upt} by computing the stationary averages by
integrating equations \eqref{x1dot} and \eqref{x2dot}.
in $d$ dimensions we rewrite the virial equation appearing in the table \ref{table3} as:
\be
p_v d L^d=-\sum_i^2  \langle \F^{ext}_i \cdot \rr_i \rangle =-\alpha^2 \langle   (\rr_1-\rr_2)^2 \rangle+
\gamma  \sum_i \langle \uu_i\cdot\rr_i \rangle
\label{virialbaiesi5bis}
\ee
The average of the operator $\OO= \sum_i\uu_i\cdot  \rr_i $ in the limit of $t\to \infty$ reads:
\bea
& &
- \frac{1}{\gamma} \sum_i \langle \F^{ext}_1 \cdot \uu_1 \rangle= 
- \frac{ \alpha^2}{\gamma} \langle (\rr_1-\rr_2) \cdot (\uu_1-\uu_2) \rangle-
\nonumber 
\\
& &
\sum_i\frac{1}{\tau}   \langle    \uu_i  \cdot \rr_i \rangle+ \sum_i \langle    \uu_i\cdot\uu_i \rangle
\label{virialbaiesib7b}
\eea
where the last term   is given by $2 d D/\tau$, and $ \langle (\rr_1-\rr_2) \cdot (\uu_1-\uu_2) \rangle=  \frac{D d}   {1+\frac{\tau}\gamma\Omega^2}  $.  After rearranging we  
find $
 \sum_i \langle    \uu_i  \cdot \rr_i \rangle= 
   d  D \Bigl(\frac{1}   {1+\frac{\tau}\gamma\omega^2} +
   \frac{1}   {1+\frac{\tau}\gamma\Omega^2}\Bigr)
$
and using the result 
$
\langle   (\rr_1-\rr_2)^2 \rangle=\frac{2d}{\Omega^2} \frac{D \gamma}{ 1+\frac{\tau}{\gamma} \Omega^2}
$
we see that the virial pressure, $p_v$, of the  GCN model coincides with formula \eqref{Upt} 
which was obtained by the UCNA method.
{ Notice that the r.h.s. of  eq. \eqref{virialbaiesi5bis} displays the characteristic structure of the pressure in active systems.
The first term is negative and contains the contribution to the pressure stemming from two-particle direct  interactions, while the second represents the combination of 
the ideal and active contributions to the pressure.}

\subsubsection{RU+GCN model: thermodynamic route}

In order to apply the "thermodynamic' route,
we  consider the following system of Markovian processes 
and recast \eqref{x1dot} and \eqref{x2dot} in terms of the collective coordinates $q,Q$:
\bea
\dot q&=&v \label{q1}\\
 \dot Q&=&V\\
 \dot v&=&-\frac{1}{\gamma}\Omega^2 v -\frac{v}{\tau}-\frac{ \Omega^2   }{\gamma\tau} q    +\frac{ D^{1/2}}{\tau}  \sqrt{2} \eta_q \\
 \dot V&=&-\frac{1}{\gamma} \omega^2 V -\frac{V}{\tau} -\frac{\omega^2}{\gamma\tau} Q   +\frac{ D^{1/2}}{\tau}  \frac{1}{\sqrt{2}} \eta_Q \, ,
 \label{vgrande}
 \eea
 By using the method illustrated in appendix \ref{thermodynamicRUGCN} we obtain the following expression
 for the probability  distribution of the coordinates $q,Q$: 
 \begin{widetext}
\be
 P_c^{config} (q,Q) = 
 \frac{1}{Z} \exp\Bigl\{- \frac{\omega^2 Q^2  (1+\frac{\tau}{\gamma} \omega^2)   
+\frac{\Omega^2}{4}q^2 (1+\frac{\tau}{\gamma} \Omega^4)}{D \gamma} \Bigr\} \, (1+\frac{\tau}{\gamma} \omega^2) (1+\frac{\tau}{\gamma} \Omega^2) \, .
\label{pconf}
\ee
\end{widetext}
The normalizing factor $Z$ is identical to the one already found in the UCNA treatment and 
also the "thermodynamic pressure" $\beta p_t= \frac{\partial }{\partial L} \ln Z$
is the same as \eqref{Upt}.
Let us remark that the above results can be easily extended to the $d$ dimensional case by substituting
$(q,Q) \to ({\bf q},{\bf Q})$ and the factor $ (1+\frac{\tau}{\gamma} \omega^2) (1+\frac{\tau}{\gamma} \Omega^2)\to 
 (1+\frac{\tau}{\gamma} \omega^2)^d (1+\frac{\tau}{\gamma} \Omega^2)^d$.

\subsubsection{ RU+GCN model:  distribution functions approach}
In order to establish the equivalence between the pressure derived by the  distribution functions approach and the virial and thermodynamic methods in the framework of the RU+GCN model
we verified that the form  $P^{config}_c$ \eqref{pconf} satisfies the balance equation \eqref{bgy}, by  using the form of the equations  \eqref{aevolution1}
 and  \eqref{aevolution2} .
  Thus we conclude 
 even in the GCN,  the pressure computed by the distribution function approach is identical  to the pressure
 of the virial method and, perhaps more interestingly, the GCN and the UCNA give identical results.   
 
 \subsubsection{Virial pressure in the ABP model}
 \label{virialABP}
{ We finally compute the pressure for the ABP model using the virial approach.}
 This is the only example  where we are able to give an explicit representation of the pressure
in ABP systems.
 Let us consider
the following equations of evolution  
\bea
\dot \qq&=&-\frac{1}{\gamma}\Omega^2 \qq  + v_0 \eq 
\label{qq}\\
 \dot \QQ&=&-\frac{1}{\gamma} \omega^2  \QQ+ v_0 \eQ
 \label{QQ} 
\eea
with
$
\eq={\bf e}_1-{\bf e}_2$ and
$\eQ=\frac{({\bf e}_1+{\bf e}_2)}{2}$.
By applying the virial formula \eqref{virialbaiesi1} 
we find
\be
p_v V d = -\alpha^2  \langle \qq^2 \rangle+ \gamma  v_0 \frac{ \langle \eq \qq+ 4 \eQ \QQ\rangle }{2}
\label{virialbaiesi8b}
\ee
and after computing  the steady-state averages featuring in the r.h.s. of eq. \eqref{virialbaiesi8b} with the help of eqs. \eqref{qq}, \eqref{QQ} and  the relations
$\langle \eq  \qq  \rangle= 2 \frac{v_0 \tau}{ 1+\frac{\tau}{\gamma}\Omega^2 } $ and
$\langle \eQ  \QQ  \rangle= \frac{1}{2} \frac{v_0 \tau}{ 1+\frac{\tau}{\gamma}\omega^2 } $ 
we  find that the virial pressure for the RU+ABP model is given by a formula identical to eq. \eqref{Upt}.
Again we notice that the first term in the r.h.s. of eq. \eqref{virialbaiesi8b} is the virial of the interparticle forces, i.e. the so-called direct interaction term of the
pressure, which has an analog in passive systems, while the second term is identified with the active pressure due to the diffusion of the 
particles and following the literature is named "swim virial" \cite{takatori2014swim}.
Such an explicit  result is valid for arbitrary $\tau$ and relies on the linearity of the forces.

\section{Conclusions}
\label{conclusions}
In this paper, we have considered models of interacting active particles, characterized by different types of stochastic drivings, corresponding to  ABP,  GCN and  UCNA dynamics. 
By focusing on the non-equilibrium steady-states of these models  we have discussed in detail  the notion of pressure, which by analogy with equilibrium systems can be derived  (in some of the cases here investigated) from the analog of the partition function, from the virial theorem or from the calculation of the stress tensor.
In particular, by considering an explicit model, that is a system of active elastic  dumbbells confined by parabolic wells, we have found that the pressure in each model is the same  independently from the differences in  their dynamical evolution laws. In the case of the elastic dumbbells, the perfect agreement among different methods and different dynamical models holds to all orders in $\tau$ and not only to first order, as predicted by the more general theory\cite{marconi2016pressure}. In the case of the UCNA, this can be understood as a consequence of the detailed balance condition which is implicitly assumed in the approximation~\cite{marconi2016heat}.
 {The GCN, in general, does not enjoy of  the detailed balance
condition but in the case of harmonic forces, such a condition is satisfied. This is the reason why in the RU model the 
 GCN  and the UCNA pressure have the same value \cite{marconi2016heat}. Finally, according to the virial method, the pressure in the  RU+ABP model is the same as the two other models as the explicit solution shows, but the two remaining methods cannot be applied. }
Let us comment that the present results regarding the pressure of a gas of underdamped active dumbbells are different from those
recently obtained by Joyeux and Bertin~\cite{joyeux2016pressure} for two reasons: 
we assumed  overdamped dynamics and
linear forces thus excluding any  average torque acting locally on the dumbbells.

\appendix

{ 
\section{Equivalence between the active parts of  the GCN and UCNA pressures}
\label{equivalence}

In the present appendix, we show the equivalence between the  active pressure contribution  obtained via the UCNA  eq. \eqref{virialbaiesivv}:
\bea
 & & \delta p^{UCNA} V d\equiv D\gamma 
 \sum_{\alpha i}  \left\langle   \Gamma^{-1}_{\alpha i,\alpha i}  \right\rangle = \nonumber \\ 
 & &
 D\gamma\left\langle Tr\Bigl[\delta_{ij} \delta_{\alpha \beta}- \frac{\tau}{\gamma }  
 \frac{\partial F^{ext}_{\alpha i} }{\partial r_{\beta j} } - 
 \frac{\tau}{\gamma }   \frac{\partial  F^{int}_{\alpha i} }  {\partial r_{\beta j}  }   \Bigr ]^{-1}\right\rangle 
 \label{b2}
 \eea
and the one via
the GCN, eq. \eqref{virialbaiesi1}:
\bea
& &
\delta p^{GCN} V d\equiv \gamma  \sum_i \langle \uu_i\cdot\rr_i \rangle = 
\nonumber \\ 
& &
N d D\gamma + \tau \left \langle \sum_i  \F^{ext}_i \cdot \uu_i + \frac{1}{2} \sum_{ij}'  \F_{ij} \cdot ( \uu_i-\uu_j)  \right\rangle .\nonumber\\
\label{b1}
\eea
 Clearly, in the noninteracting case and with $\tau=0$ the two formulas give the same active  pressure $N D\gamma/V$.
 Hereafter, we show that  to first order in $\tau$ \eqref{b2} and  \eqref{b1} give the same contribution to the total pressure.
 By comparing the $\langle \uu_i\cdot\rr_i \rangle$ term in \eqref{b1} with $D \sum_{\alpha }  \sum_{ i}\langle   \Gamma^{-1}_{\alpha i,\alpha i}  \rangle$  term in \eqref{b2} and with the help  of 
 Novikov's theorem  \cite{novikov1965functionals} we shall  prove the equivalence. 
  Let us consider the stochastic equation for the GCN model:
 \begin{eqnarray}
\frac{d \bar x_k(t)}{dt}\!\!&=&\! \frac{F_k( x)}{\gamma} +  u_k(t), 
\label{has1}
\end{eqnarray}
where the index $k$ stands for $(\alpha,i)$ and $x_k=r_{\alpha i}$.
 The Novikov theorem states that for a Gaussian process (with or without memory) \cite{fox1983correlation,rein2016applicability}
  the average
$\langle u_m(t) x_k(t)\rangle $ is given by:
\begin{eqnarray}
\langle u_m(t) \Phi(u) \rangle
&=&\int_0^t \:dt' c_{mm}(t,t')  
\left< \frac{\delta \Phi(u)}{\delta u_n} \right>,
\label{has5}
\end{eqnarray}
where $\Phi(\{u\})$ denotes a function of
the $\{u\}$ , and $c_{mn}(t,t')$ is the time correlation function 
 $ c_{mn}(t,t')=\langle u_m(t) u_n(t') \rangle=\delta_{mn}\frac{D}{\tau}\exp(-\frac{|t-t'|}{\tau}).
  $
 In our case eq. \eqref{has5} becomes
 \be
 \langle x_k(t) u_m(t)\rangle= \int_0^t dt' c_{mm}(t,t') \langle \frac{\delta x_k(t)}{\delta u_m(t')} \rangle
 \label{xut}
 \ee
 After integrating Eq. \eqref{has1}, we get
\begin{eqnarray}
\bar x_k(t)=x_k(0)+\int_0^t \:ds \:[\frac{F_k(\bar x(s))}{\gamma}+ \, u_k(s)].
\label{has8}
\end{eqnarray}
The functional derivative of $\bar x_k(t)$  with respect to
$u_m(t')$ becomes
\begin{eqnarray}
& & \frac{\delta \bar x_k(t)}{\delta u_m(t')} =
\nonumber \\
& & 
\delta_{km}\theta(t-t')+ \int_{t'}^{t}\:ds \: \frac{1}{\gamma} \frac{\partial F_k(\{\bar x(s)\})}{\partial x_n(s)}
\frac{\delta \bar x_n(s)}{\delta u_m(t')}.
\label{has9}
\end{eqnarray}
and the derivative of Eq. \eqref{has9} with respect to $t$ is given by
\begin{eqnarray}
& &
\frac{\partial}{\partial t} 
\left[\frac{\delta \bar x_k(t)}{\delta u_m(t')} \right]
= \nonumber \\
& & \delta_{km}\delta(t-t')    +  \frac{1}{\gamma} \frac{\partial F_k(\{\bar x(t) \})}{\partial x_n(t)}
\frac{\delta \bar x_n(t)}{\delta u_m(t')}.
\label{has10}
\end{eqnarray}
Define, now,   $H_{kn}(t) = - \frac{1}{\gamma}  \frac{\partial F_k(\{x(t)\})}{\partial x_n(t)} $ 
and rewrite \eqref{has10} as:
\begin{eqnarray}
\frac{\partial}{\partial t} 
\left[\frac{\delta \bar x_n(t)}{\delta u_m(t')} \right]
&=&  \delta_{mn}  \delta(t-t')    -  H_{kn}(t) 
\frac{\delta \bar x_n(t)}{\delta u_m(t')}, 
\label{has11}
\end{eqnarray}
whose  formal solution  with the initial condition 
\begin{eqnarray}
\left[\frac{\delta\bar x_n(t)}{\delta u_m(t')} \right]_{t=t'}&=&\delta_{mn},
\label{has12}
\end{eqnarray}
 is given (for $t>t'$) by
\begin{eqnarray}
\frac{\delta \bar x_n(t)}{\delta u_m(t')}&=&   \theta(t-t')
 \exp\left( -\int_{t'}^{t}\:ds \:  {\bf H}(s)\right)_{nm} .
\label{has13}
\end{eqnarray}

{
Combining eq. \eqref{xut} with eq. \eqref{has13}, we get
\begin{eqnarray}
& &
\langle x_k(t) u_m(t) \rangle
= \nonumber \\
& &
\int_0^t dt'  \theta(t-t')
 c_{mm}(t,t') \left< \exp\left(-\int_{t'}^{t} \:ds   {\bf H}(s) \right)_{km}  \right>. \nonumber\\
\label{has14} 
\end{eqnarray}
By a change variable $z\equiv (t-t')/\tau$ and substitution of the explicit expression of $c_{mm}(t,t')$, we obtain
\begin{eqnarray}
& &
\langle x_k(t) u_m(t) \rangle
= \nonumber \\
& &
D \int_0^{t/\tau} dz  \exp(-z) 
  \left< \exp\left(-\int^{t}_{t-\tau z} \:ds   {\bf H}(s) \right)_{km} \right> . \nonumber\\
\label{has14b}
\end{eqnarray}
By  taking the small $\tau$ limit, i.e. assuming that $\bar x_n$ does not vary significantly over the 
correlation time of the noise, we can express the integral in the exponent as
$$
\int^{t}_{t-\tau z} \:ds   {\bf H}(s) \approx \tau z {\bf H}(t) \,  -\frac{1}{2} \frac{d {\bf H}(t)}{d t} \tau^2 z^2+\dots
$$
Neglecting the quadratic term proportional to $\tau^2 z^2$ the explicit form of \eqref{xut} reads
\bea
\langle x_m(t) u_m(t) \rangle
 \approx   
D \Bigl\langle \Bigl(  {\bf I}+\tau {\bf H} \Bigr)^{-1}_{mm}  \Bigr\rangle
\eea
Finally, going back to the physical variables of interest we obtain the result}
\bea
& &
\gamma  \sum_i \langle \uu_i\cdot\rr_i \rangle=
\gamma \sum_k  \langle u_k(t)  x_k(t) \rangle
=\nonumber \\
& &
D\gamma \Bigr\langle \langle Tr\Bigl[\delta_{ij} \delta_{\alpha \beta}- \frac{\tau}{\gamma }  
 \frac{\partial F^{ext}_{\alpha i} }{\partial r_{\beta j} }- \frac{\tau}{\gamma }   \frac{\partial  F^{int}_{\alpha i} }  {\partial r_{\beta j}  }   \Bigr ]^{-1}
 \Bigr\rangle 
  \eea
 Thus, we have shown 
that the two expression for the active pressure, in GCN and UCNA models, coincide to first order in $\tau$. 
{ Unfortunately, attempts to include the ABP model in this analysis are hampered by the non Gaussian nature
of the noise distribution associated with the corresponding $u_n(t)$. Only in the case of linear forces it is possible to derive an explicit result for the pressure.}
}

{
\section{Mechanical pressure in the RU UCNA}
\label{mechanicalRU} 

 The exact steady-state distribution function for $N=2$ particles takes the simple form  
 \bea
 & &
  -D\gamma \sum_\beta \sum_{j=1,2} \frac{\partial }{\partial r_{\beta j}}[  \Gamma^{-1}_{\alpha 1,\beta j}(\rr_1,\rr_2) P_2(\rr_1,\rr_2)]= 
  \nonumber \\
  & &
   P_2(\rr_1,\rr_2) \Bigl(  \frac{\partial u(\rr_1)}{\partial r_{\alpha 1}} +     \frac{ \partial w(\rr_1-\rr_2) }{ \partial r_{\alpha 1}}   \Bigl) .
\label{bgy}
\eea  
Integrating both sides of equation \eqref{bgy} with respect to $\rr_2$ and recalling that $\bGamma$ is constant  for the oscillator problem we obtain
 \bea
 & &
 P^{(1)}_2(\rr_1) F_{\alpha}^{ext}(\rr_1)
 \nonumber \\
 & &
  =D\gamma  \Gamma^{-1}_{\alpha 1,\alpha 1} \frac{\partial }{\partial r_{\alpha 1} }P^{(1)}_2(\rr_1)  
  +    \int d\rr_2 P_2(\rr_1,\rr_2)  \frac{ \partial w(\rr_1-\rr_2) }{ \partial r_{\alpha 1}}   \nonumber \\
  \label{forcef}
\eea  
where $P^{(1)}_2(\rr_1)\equiv \int d\rr_2   P_2(\rr_1,\rr_2) $ is the marginalized one particle distribution function
and $\F^{ext}(\rr)= -\frac{\partial u(\rr)}{\partial \rr} $.
Following ref. \cite{marconi2016pressure} we now use eq. \eqref{forcef}  to derive an  expression for the pressure
in terms of the work of deformation, $\delta W_F$, necessary to produce a change of volume $\delta V$. We obtain  $\delta W_F$  
by  multiplying the l.h.s. of eq.  \eqref{forcef} by an infinitesimal  displacement $\delta {\bf s} (\rr)=\lambda \rr$
and integrating over the volume:
\be
\delta W_F=  \int d^d \rr \, \rho(\rr)  {\bf F^{ext}} (\rr)\cdot \delta {\bf s} (\rr)
  \label{forcefb}
\ee
where we introduced the particle density via $\rho(\rr)=2 P^{(1)}_2(\rr)$. 
 Using eq. \eqref{forcef}   we get
\be
\delta W_F=- \lambda d D\gamma 
\Bigl( 
\frac{1}   {1+\frac{\tau}\gamma\omega^2}+ \frac{ \omega^2 }{ \Omega^2} 
\frac{1}   {1+\frac{\tau}\gamma\Omega^2} \Bigr )
\label{wf1} 
\ee
On the other hand, one can evaluate the work of deformation 
as the integral over the volume  of the
trace of the product of  the pressure 
tensor $p_{\alpha\beta}(\rr) $ times the strain tensor, $\delta \epsilon_{\alpha\beta}$, associated with
the local displacement  $\delta {\bf s(\rr)}$ of the fluid.
With the help of the explicit formula
$\delta \epsilon_{\alpha\beta}=\frac{1}{2}( \frac{\partial \delta s_\alpha}{\partial r_\beta}+  \frac{\partial \delta s_\beta}{\partial r_\alpha})=
\lambda \delta_{\alpha\beta} $  such a  work can be calculated as:
\be
\delta W_p=-  \int d\rr  \sum_{\alpha\beta}p_{\alpha\beta}(\rr) \delta \epsilon_{\alpha\beta} (\rr) =-\lambda d  L^d  \, p_V
\label{wp1}
\ee
where $ p_V$ is the volume averaged pressure tensor.
Since  $\delta W_F$ and $\delta W_p$ must be equal,
by comparing the  r.h.s. of eq.\eqref{wf1}  and \eqref{wp1} we see that  $p_V$ 
is identical to the virial pressure of formula  \eqref{Upt}.

 \section{Thermodynamic pressure in the Riddell-Uhlenbeck GCN model} 
 \label{thermodynamicRUGCN}
We write the Kramers equation associated with  eqs.~\eqref{q1}-\eqref{vgrande} for the joint distribution function, $P_c$, of the collective variables:
\begin{widetext}
\bea
\frac{\partial  P_c (q,v,Q,V,t)}{\partial t }&+&
v \frac{\partial } {\partial q}  P_c-\frac{1} {\tau}  (1+\frac{\tau} {\gamma} \Omega^2)  \frac{\partial }{\partial v} v P_c
    -\frac{1}{\tau \gamma}\Omega^2 q  \frac{\partial }{\partial v} P_c  \nonumber \\
    &+&
V \frac{\partial } {\partial Q}  P_c -\frac{1} {\tau}  (1+\frac{\tau} {\gamma} \omega^2)  \frac{\partial }{\partial V} V P_c 
    -\frac{1}{\tau \gamma}\omega^2 Q \frac{\partial }{\partial V} P_c
  =\frac{ 2 D}{\tau^2}   \frac{\partial^2 P_c}{\partial v^2} +
    \frac{  D}{ 2 \tau^2}   \frac{\partial^2 P_c}{\partial V^2} 
\label{poevolution2}
\eea
\end{widetext}
and look for a time-independent solution whose momentum currents $j_v(q,Q)\equiv\int dv dV v P_c(q,v,Q,V)$ 
and  $j_V(q,Q)\equiv\int dv dV V P_c(q,v,Q,V)$ vanish  for all values of $q$ and $Q$ under the following form \cite{marconi2016velocity}:
\bea
& &
P_c^{st}(q,Q,v,V) = 
\nonumber \\
& &
P_c^{config} (q,Q)\exp\Bigl\{- \frac{ \tau}{D} \Bigl[(1+\frac{\tau}{\gamma}\omega^2)  V^2+\frac{1}{4}   (1+ \frac{\tau}{\gamma}\Omega^2  )
v^2\Bigr] \Bigr\} . \nonumber \\
\label{p2colored}
\eea
Substituting  \eqref{p2colored}  into eq. \eqref{poevolution2},
multiplying the latter by  $v$ and by $V$, integrating over velocities and imposing that  currents vanish
we obtain the following two equations which  determine  $ P_c^{config} (q,Q)$: 
 \be
  \frac{1} {1+\frac{\tau} {\gamma} \Omega^2 }  \frac{\partial } {\partial q}  
    P_c^{config} (q,Q) +    \frac{1}{ 2 D\gamma}\Omega^2 q     P_c^{config} (q,Q) =0  \\
\label{aevolution1}
\ee
\be 
 \frac{1} {1+\frac{\tau} {\gamma} \omega^2 }   \frac{\partial } {\partial Q}    P_c^{config} (q,Q)  +    \frac{2}{D\gamma}\omega^2 Q    P_c^{config} (q,Q) =0 
\label{aevolution2}
\ee
which are compatible with the solution given by formula \eqref{pconf}.
}

\section{Pressure of independent particles driven by two state noise in a  one dimensional harmonic well}
In this appendix, we revisit the treatment Hanggi and Jung \cite{hanggi1995colored} to provide one more pressure formula 
relative to a model of independent active particles confined
in a parabolic one-dimensional well. The interest stems from the fact that the model comprises a dichotomic noise and the
steady-state  density distribution is peaked  near the boundaries, instead of being a Gaussian, but the resulting  pressure 
formula is identical to the formulae reported in the main text in the limit $\alpha=0$ .
Let us consider the one-dimensional run-and-tumble  model of Schnitzer~\cite{schnitzer1993theory} described by
\be
\dot x_{ i} =  -  \frac{\omega^2}{\gamma} x_{ i}+v_0 \psi_i(t)
\label{stochasticxhanngi}
\ee
where the dichotomic noise is
$\psi_i(t)= (-1)^{n_i(t)}$
and $n(t)$ is a Poisson process with parameter $\lambda$, such that 
$
P(n(t)=m)= \frac{(\lambda t)^m}{m!}\, \exp(-\lambda t) $, 
so that for $\psi_i(0)=1$ the average is
$\langle \psi_i(t) \rangle= \exp(-2 \lambda t) $
and the time-correlator is:
$\langle \psi_i(t) \psi_j(s) \rangle= \delta_{ij}\exp(-2 \lambda |t-s|) $.
The corresponding stationary distribution of positions  for $x^2\leq \gamma v_0/\omega^2$ is:
\be
Prob(x)=\frac{1}{Z}\Bigl(v_0^2-\frac{\omega^4 x^2}{\gamma^2}
 \Bigr)^{\frac{\lambda\gamma}{\omega^2}-1}
 \ee
 and zero otherwise and $Z$ is a normalizing factor $Z   = \int dx \, \left[ 1 - (\frac{\mu \omega^2}{v})^2 x^2 \right]^\beta$.
 We, now, consider
 \be
 \langle x^2 \rangle =   (\frac{\gamma v_0}  {\omega^2 })^2    \frac{ \int_{-1}^1 dy y^2  \Bigl(1-y^2 \Bigr)^\alpha }{\int_{-1}^1 dy   \Bigl(1-y^2 \Bigr)^\alpha 
}=    (\frac{\gamma v_0}  {\omega^2 })^2  R(\alpha)
\ee 
 with $y= \frac{\omega^2 }{\gamma v_0} x$ and $\alpha={\frac{\lambda\gamma}{\omega^2}-1}$. 
 The integral exists if $\alpha>-1$.
 For $\alpha$ integer or semi-integer we use the following exact formula
 $R(\alpha)=\frac{1}{3+2 \alpha}
 $.
 Finally, we  compute the virial, $\langle Fx\rangle=-p_v L $, and obtain:
 \be
 p_v= \frac{N}{L} \frac{D \gamma}{1+\frac{\omega^2 \tau}{\gamma} } .
 \ee
 We may conclude that in this linear problem, as far as the pressure is concerned, the only thing which matters is the character of the pair correlations
 that is the exponentially decaying character.
  In fact, such a result follows from
\bea
& &
\lim_{t\to\infty}  \langle x_i(t)  x_i(t)\rangle =  
\nonumber 
\\
& &
v_0^2 e^{-\frac{2\omega^2}{\gamma} t} \int_0^t dt_1  \int_0^t dt_2 e^{\frac{\omega^2 }{\gamma}( t_1+t_2)} e^{ -2\lambda| t_1-t_2|}
\eea
\be
\lim_{t\to\infty} \langle x_i(t)  x_i(t)\rangle=\frac{v_0^2\gamma}{\omega^2}  \frac{1} {2\lambda+\frac{\omega^2}{\gamma}}=\frac{1}{\omega^2} \frac{D \gamma}{1+\frac{\omega^2 \tau}{\gamma} }
\ee

\section*{Acknowledgments}
C. Maggi acknowledges support from the European Research Council under the European Union's Seventh Framework programme
(FP7/2007-2013)/ERC Grant agreement  No. 307940.
MP was supported by the Simons Foundation  
Targeted Grant in the Mathematical Modeling of Living Systems Number: 342354 and by the Syracuse Soft Matter Program.

%
\end{document}